# DGNN-Booster: A Generic FPGA Accelerator Framework For Dynamic Graph Neural Network Inference

Hanqiu Chen
Electrical and Computer Engineering
Georgia Institute of Technology
Atlanta, USA
hchen799@gatech.edu

Cong Hao
Electrical and Computer Engineering
Georgia Institute of Technology
Atlanta, USA
callie.hao@ece.gatech.edu

*Abstract*—Dynamic Graph Neural Networks (DGNNs) are becoming increasingly popular due to their effectiveness in analyzing and predicting the evolution of complex interconnected graph-based systems. However, hardware deployment of DGNNs still remains a challenge. First, DGNNs do not fully utilize hardware resources because temporal data dependencies cause low hardware parallelism. Additionally, there is currently a lack of generic DGNN hardware accelerator frameworks, and existing GNN accelerator frameworks have limited ability to handle dynamic graphs with changing topologies and node features. To address the aforementioned challenges, in this paper, we propose DGNN-Booster, which is a novel Field-Programmable Gate Array (FPGA) accelerator framework for real-time DGNN inference using High-Level Synthesis (HLS). It includes two different FPGA accelerator designs with different dataflows that can support the most widely used DGNNs. We showcase the effectiveness of our designs by implementing and evaluating two representative DGNN models on ZCU102 board and measuring the end-to-end performance. The experiment results demonstrate that DGNN-Booster can achieve a speedup of up to $5.6\times$ compared to the CPU baseline (6226R), $8.4\times$ compared to the GPU baseline (A6000) and $2.1\times$ compared to the FPGA baseline without applying optimizations proposed in this paper. Moreover, DGNN-Booster can achieve over $100\times$ and over $1000\times$ runtime energy efficiency than the CPU and GPU baseline respectively. Our implementation code and on-board measurements are publicly available at https://github.com/sharc-lab/DGNN-Booster.

## I. INTRODUCTION

Graph Neural Networks (GNNs) are powerful tools for capturing relationships within graph-structured data and can be applied in a wide range of domains, including recommendation systems [1], drug discovery [2], fraud detection [3] and traffic prediction [4]. In real-world applications, DGNNs have several advantages over traditional GNNs. They can capture temporal dependencies between the nodes and edges in a graph and thus achieve better performance in temporal-related tasks such as traffic pattern prediction [5] and stock price forecasting [6]. Additionally, DGNNs are highly flexible by combining different types of GNNs for spatial encoding and Recurrent Neural Networks (RNNs) for temporal encoding, resulting in improved performance.

While different types of DGNNs have seen success in software [7]–[9], challenges remain in their hardware deployment: (1) *Low parallelism*. It is hard to parallelize the computation on hardware due to temporal data dependencies between graphs at different times. (2) *Large memory consumption and frequent memory access*. The time-evolving graph embeddings leads to large memory consumption and frequent data transfer between on-chip and off-chip memory. (3) *High energy consumption*: DGNNs have high energy consumption due to computation-intensive matrix multiplications and complex mathematical operations.

To address the aforementioned challenges, in this paper, we propose **DGNN-Booster**, which is a generic FPGA accelerator framework for DGNNs that achieves high-speed and low energy consumption on-board inference and can be applied to various popular DGNNs. Our contributions can be summarized as follows:

1) **Generic and open-source.** DGNN-Booster is a model-generic framework, developed using High-Level Synthesis (HLS) for ease of use. It has modularized processing elements (PEs) for GNN and RNN and supports multiple types of GNNs and RNNs. It's publicly available, with on-board measurement and end-to-end functionality verified by crosschecking with PyTorch code.
2) **Hardware efficient.** DGNN-Booster has multi-level parallelism with hardware architecture optimizations, aiming to deliver real-time performance with lower energy consumption compared to CPU and GPU. Different graphs at different time steps can be streamed in consecutively and processed on-the-fly.
3) **Two accelerator designs with different dataflows.** DGNN-Booster has two designs that support different dataflows between GNN and RNN. DGNN-Booster V1 overlaps them in adjacent time steps and DGNN-Booster V2 connects them in data streaming within one time step. Both designs feature data streaming inside RNN for increased parallelism and lower on-chip memory consumption.
4) **On-board evaluation.** We verify DGNN-Booster on Xilinx ZCU102 FPGA using two popular temporal graph datasets with varying graph sizes at different time steps based on two representative DGNN models. We also do an ablation study to demonstrate the effectiveness of our multi-level parallelism design.

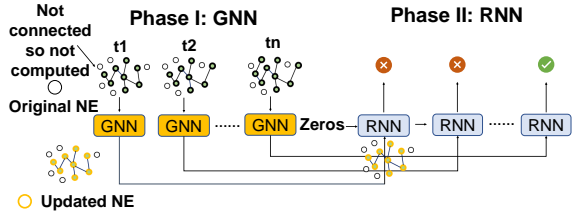

Fig. 1: A high-level overview of CPU and GPU implementation dataflow of stacked DGNNs. The output from GNN at different time steps will be fed into RNN in a sequential manner. Only the output from the RNN in the last time step will be used for the following computation. (NE: node embedding)

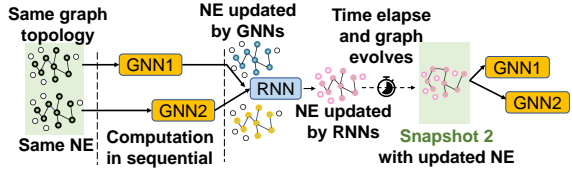

Fig. 2: A high-level overview of CPU and GPU implementation dataflow of integrated DGNNs. The output from RNN in the last time step will be used as the input of GNN in the next time step. Two GNNs are in sequential. The GNNs and RNN are also computed in a sequential manner. (NE: node embedding)

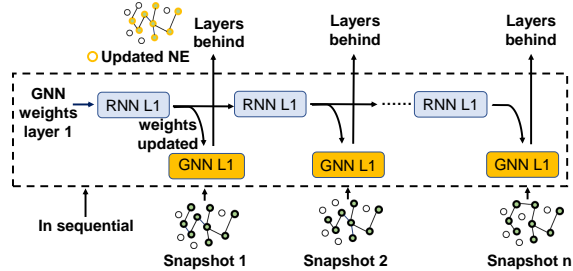

Fig. 3: A high-level overview of CPU and GPU implementation dataflow of weights-evolved DGNNs. Weights are evolved by RNN and used by GNN in a sequential manner. (NE: node embedding)

## II. BACKGROUND ABOUT DGNNS

Dynamic Graph Neural Networks (DGNNs) are GNNs designed for dynamic graph structures and features. According to a recent survey [10], DGNNs can be classified into two different categories: discrete-time DGNNs and continuous-time DGNNs. DGNN-Booster supports discrete-time DGNNs, which use a set of ordered graphs (snapshots) to represent dynamic graphs.

$$DG = \{G^1, G^2, ..., G^T\} \quad (1)$$

where T is the number of snapshots. This discrete-time representation of dynamic graphs enables the use of traditional static GNNs for spatial information encoding and RNNs for temporal information encoding.

We identify three types of discrete-time DGNNs based on the dataflow relationship between GNN and RNN, with a summary in Table I.

- **Stacked DGNNs.** It is the most straightforward way to model a discrete-time dynamic graph. GNN encodes them as time-series information and feeds them into RNN. This process can be represented as:

$$\begin{aligned} X^1, X^2, ..., X^t &= \text{GNN}(G^1, G^2, ..., G^t) \\ O^{t+1} &= \text{RNN}(X^1, X^2, ..., X^t) \end{aligned} \quad (2)$$

where $G^t$ is the node embedding of the snapshot at time $t$, $X^t$ is the updated node embedding by GNN at time $t$, $O^{t+1}$ is the output of RNN for time $t+1$. The high-level dataflow diagram of this type of DGNN is shown in Fig. 1.

- **Integrated DGNNs.** This type of discrete-time graph encoding combines GNN and RNN together within one time step by replacing matrix multiplications in RNN with graph-related operations, such as graph convolution. It can be expressed as the following equations:

$$\begin{aligned} X_1^t &= \text{GNN1}(G^t) \\ X_2^t &= \text{GNN2}(G^t) \\ G^{t+1} &= \text{RNN}(X_1^t, X_2^t) \end{aligned} \quad (3)$$

where $G^t$ represents the node embedding of the snapshot at time $t$. $X_1^t$ and $X_2^t$ are two different updated node embeddings by GNN1 and GNN2 with different weights. The high-level dataflow diagram of this type of DGNN is shown in Fig. 2.

- **Weights-evolved DGNNs.** The dataflow between GNN and RNN of this type of DGNN is similar to Stacked DGNNs. The difference is in what is evolved by RNN. Different from stacked DGNN, where node embeddings updated by GNN are evolved by RNN, the weights of GNN are evolved by RNN. It can be expressed as the following equations:

$$\begin{aligned} W^t &= \text{RNN}(W^{t-1}) \\ O^t &= \text{GNN}(W^t, G^t) \end{aligned} \quad (4)$$

where $W^t$ is the weight of GNN at time $t$, $G^t$ is the node embedding at time $t$, $O^t$ is the output of GNN at time $t$. The high-level dataflow diagram of this type of DGNN is shown in Fig. 3.

## III. MOTIVATIONS AND INNOVATIONS

### A. Related Works and motivations

There are some recent developments in DGNN hardware accelerators. Zhou et al. [19] perform model-architecture co-design on memory-based Temporal Graph Neural Networks. Cambricon-G [20] is the first hardware accelerator aiming to exploit more opportunities for data reuse using multidimensional multilevel tiling. Chakaravarthy et al. [21] finish the first scaling study on DGNNs by designing a multi-GPU DGNN training system. DynaGraph [22] and TGL [23] are another two high-performance DGNN training frameworks on GPU focusing on spatial and temporal knowledge unifying and simple user configuration.

TABLE I: A summary of different types of discrete-time DGNNs and their dataflow types. We also showcase which accelerator design in DGNN-Booster can be applied.

| DGNN type | Related works | Dataflow type | DGNN Booster V1 | DGNN Booster V2 |
|---|---|---|---|---|
| Stacked DGNN | GCRN-M1 [11], RgCNN [12] WD-GCN [13], DyGGNN [14] | • Data dependencies between GNN and RNN within one time step.<br>• Independent GNN at different time steps. | ✓ | ✓ |
| Integrated DGNN | GCRN-M2 [11], GC-LSTM [15] LRGCN [16], RE-Net [17] | • Data dependencies between GNN and RNN in adjacent time steps.<br>• Dependent GNN at different time steps. | ✗ | ✓ |
| Weights-evolved DGNN | EvolveGCN [18] | • Weights for GNN are evolved by RNN.<br>• Independent GNN at different time steps. | ✓ | ✗ |

However, there still remain some challenges on DGNN hardware deployment. ❶ **High energy consumption and low computation resource utilization.** Previous works primarily focus on deploying DGNNs on GPUs. However, these designs suffer from high energy consumption and low computation resource utilization because of temporal data dependencies. ❷ **Lack of parallelism between GNN and RNN.** Previous research focuses on treating GNN and RNN as separate parts, which limits parallelism. ❸ **Lack of integrating GNN and RNN optimizations together into a single system.** Previous research usually optimizes GNN and RNN individually, which limits achieving optimal hardware efficiency.

### B. Innovations

Motivated by these challenges, we propose DGNN-Booster to achieve high-speed and low energy consumption DGNN inference on FPGA. Our design can cover most DGNN types shown in Table. I. It has several advantages over previous DGNN accelerator designs:

- **Better hardware performance with high flexibility.** DGNN-Booster has lower latency and energy consumption than CPU and GPU. Its modulized design of GNN and RNN using High-level Synthesis (HLS) allows for easy integration of different GNNs and RNNs. Additionally, GNN is implemented using the message passing mechanism, and we emphasize DGNN-Booster's support for edge embeddings, which are not considered by existing DGNN accelerators but are widely used in most GNN models.
- **Multi-level parallelism.** There are two levels of parallelism in our design. In higher-level parallelism, we parallelize GNN and RNN in adjacent time steps in DGNN-Booster V1 while parallelizing GNN and RNN within one time step in V2. The DGNN type supported is shown in Table I. Moreover, we overlap graph loading with GNN inference in V1. In lower-level parallelism, we implement GNN using message passing mechanism based on GenGNN [24]. The message passing and node transformation are in streaming in DGNN-Booster V2. Besides, we implement data streaming for different stages inside RNN in both accelerator designs.
- **Harware efficient architecture design.** We propose a task scheduling scheme to allocate the most suitable tasks for CPU and FPGA. Besides, we implement graph renumbering and format transformation to make our design more hardware efficient. Additionally, we utilize different types of RAMs on-chip to achieve memory efficiency.

## IV. HARDWARE ARCHITECTURE DESIGN

DGNN-Booster is developed based on a CPU-FPGA heterogeneous platform, where a host program loads weights and node features to DRAM and does graph preprocessing. After graph preprocessing finishes, FPGA loads prepared data to on-chip buffers via PCIe. The FPGA accelerator is optimized for various DGNNs with customized IPs designed using HLS, which contain parallel processing elements (PEs) for concurrent inference of GNN and RNN to achieve optimal hardware performance.

### A. Graph preprocessing and data communication

The input graphs to DGNN-Booster are in the coordinate (COO) format, which is the most widely used format in dynamic graph datasets. In COO format, edges are stored in an arbitrarily ordered list, where each list entry consists of the source node, the destination node, the data and the time associated with the edge. The host program is responsible for slicing the large input graph into small snapshots in the order of time based on the time splitter we choose. The time splitter should be set appropriately to ensure that the size of each snapshot is not too large or too small. During the snapshot generation, the CPU will also calculate the number of nodes edges of each snapshot.

The data communication modes of the weights and snapshot information are different. The weights are shared between different time steps, so the overhead of weight loading is a one-time cost before the computation on FPGA starts. The edge list, node embedding, edge embedding and the number of nodes and edges of each snapshot are sent from DRAM to on-chip buffers in the order of time waiting for computation on FPGA. Since the on-chip memory resources are very limited on FPGA, it is impossible to transmit the information of snapshots with different node and edge features at different time steps all at once to the on-chip buffers. As a result, only the information of the snapshot to be processed in the next time step will be sent to on-chip buffers.

### B. Graph renumbering and format transformation

During FPGA runtime, only a snapshot is stored in on-chip buffers. To ensure the correct data access to the on-chip buffer,

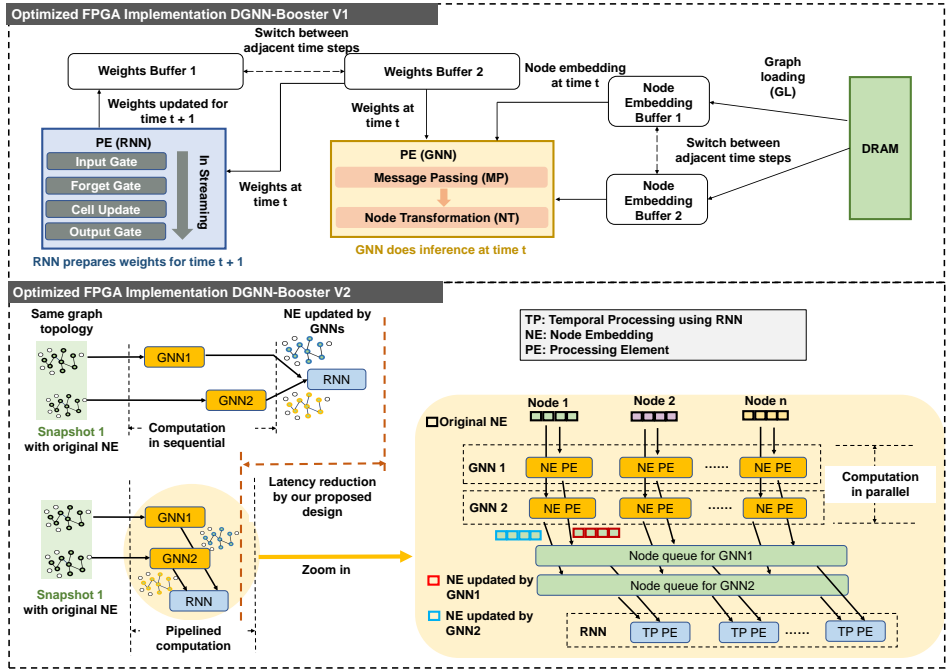

Fig. 4: Optimized FPGA implementation of DGNN-Booster V1 based on EvolveGCN and V2 based on GCRN-M2. DGNN-Booster V1 overlaps the computation of GNN and RNN in adjacent time steps by using ping-pong buffers to store the weights. Graph loading is overlapped with GNN inference using the same way. DGNN-Booster V2 overlaps the computation of GNN and RNN within one time step by utilizing node queues implemented with FIFOs.

we need to know the raw index of each node in the large raw graph and its corresponding address in the BRAM. The host program will generate a renumbering table for each snapshot to take a record of the node index renumbering information. During inference, processing units (PEs) will first check this renumbering table to do node index transformation and then correctly fetch data in the on-chip buffer. The renumbering table will also guide the FPGA to correctly fetch data from DRAM and write back.

Although the COO format is convenient for graph producers and is usually the raw data format in real-life applications, it is not hardware-friendly because finding the neighborhood nodes contains irregular computation patterns and usually brings a large overhead on FPGA. Instead of using COO format, we use compressed sparse row (CSR) format or compressed sparse column (CSC) format for GNN inference by designing a converter on FPGA for format transformation.

Graph renumbering and format transformation together will ensure the data of the snapshot is stored in a continuous space in on-chip buffers, which avoids irregular on-chip memory access.

### C. Multi-level parallelism

The techniques used to achieve multi-level parallelism are different in DGNN-Booster V1 and V2.

*1) DGNN-Booster V1:* DGNN-Booster V1 is proposed to reduce the hardware overhead caused by sequential RNN and GNN computation in stacked DGNN and weighs-evolved DGNN, whose CPU and GPU dataflows are shown in Fig. 1 and Fig. 3 respectively. This design makes GNN and RNN in adjacent time steps in parallel. Detailed FPGA implementation of DGNN-Booster V1 is shown in Fig. 4.

- **Ping-pong buffers and data-streaming FIFOs in RNN.** We avoid data conflict by using two pairs of ping-pong buffers for weights and node embeddings. As shown in Fig. 4, GNN can read the weights from buffer 2 while RNN can update the weights for the next time step and store the results in buffer 1 at the same time. Similarly, the ping-pong buffers for node embeddings allow for parallel data loading with GNN inference. We also utilize FIFOs (first-in first-out) to connect different computation stages inside RNN so that these stages can be pipelined at the node level to further reduce the latency of RNN.
- **Execution flow.** The inference process within one time step can be divided into four separate parts: graph loading (GL), message passing (MP), node transformation (NT) and RNN. Among them, MP must wait for the result from GL and NT must wait for MP and RNN. To achieve the optimal performance, we schedule RNN in $t+1$ with MP in $t$ parallel and GL in $t+1$ with NT in $t$ in parallel. This is because MP and RNN are two relatively more computation-intensive modules than GL and NT, and scheduling in this scheme can avoid workload imbalance.

*2) DGNN-Booster V2:* DGNN-Booster V2 is proposed to reduce the hardware overhead caused by sequential RNN and

GNN computation in stacked DGNN and integrated DGNN, whose CPU and GPU dataflows are shown in Fig. 1 and Fig. 2 respectively. This design makes GNN and RNN within the same time step in parallel. The high-level dataflow and FPGA architecture of DGNN-Booster V2 is shown in Fig. 4.

- **Node queues.** The node queues are implemented using FIFOs to overlap GNN and RNN computation. When the GNN finishes the updating of one node embedding, it will send the updated node embedding to the node queue. The PEs which execute the RNN will fetch node embeddings from the node queue in the same order. At the same time, GNN will be triggered to fetch new node embeddings. As a result, GNN and RNN can work in parallel on different nodes.
- **Execution flow.** In this design, the message passing and node transformation in GNN are in data streaming. Furthermore, different stages in RNN are in data streaming. Combining these optimizations with node queues connected between GNN and RNN, we achieve node-level pipelining end-to-end.

### D. Task scheduling on CPU-FPGA heterogeneous platform

To better utilize the advantages of both CPU and FPGA, we propose a task scheduling scheme. CPU offers generality over a wide range of tasks while FPGA provides high peak performance on tasks with simple computation patterns. In order to achieve optimal hardware performance, we schedule graph preprocessing and renumbering to CPU. The graph format transformation, GNN and RNN inference are scheduled to the FPGA. This is because graph preprocessing and renumber table generation need complex control flows and irregular and frequent memory access but with low computation intensity. GNN and RNN inference has many matrix multiplications, which are computation-intensive but with simple computation patterns.

### E. On-chip buffer design

There are two types of RAMs on FPGA. LUTRAM is built out of LUTs, which is more efficient for small RAMs as it can be created to fit many different sizes. For BRAM, the minimum memory size is 18KB. If any of the memory space is one BRAM is unused, it cannot be used for something else and thus wasted. DGNN-Booster contains fine-grained pipelining inside GNN and RNN. In this case, weight buffers will be partitioned into so many small RAMs, and it is a huge waste of on-chip memory resources if we store the weights in BRAM. As a result, weights are allocated to LUTRAMs. On the contrary, since the data size of node embedding and edge embedding is larger than weights, and needed to be stored in a continuous space on-chip, so we allocate them to BRAM. Our on-chip buffer design demonstrates memory efficiency and supports relatively larger snapshots to be stored on-chip to avoid frequent data exchange between the host and FPGA during the message-passing stage.

## V. EXPERIMENTS

### A. Implementation Details

We deploy DGNN-Booster using High-Level Synthesis (HLS) by Vitis HLS and Vivado for the Xilinx ZCU102 FPGA development board, as shown in Fig. 5, whose available resources are shown in Table II, targeting at 100MHz clock frequency. We choose EvolveGCN [18] as the base model for DGNN-Booster V1 and GCRN-M2 [11] for V2. Both use GCN [25] as GNN. GRU [26] and LSTM [27] are used as RNN in EvolveGCN and GCRN-M2 respectively. Both of the weights and graph embeddings are in 32-bit floating point.

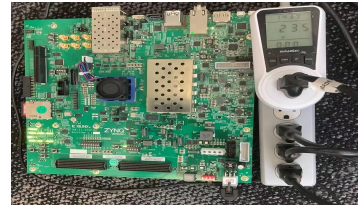

Fig. 5: On-board implementation with power measurement.

TABLE II: Resource utilization on Xilinx ZCU102 FPGA. The clock frequency is 100MHz. Results are reported by Vivado post-implementation report.

| Model | LUT | LUTRAM | FF | BRAM | DSP |
|---|---|---|---|---|---|
| **Available** | 274,080 | 144,000 | 548,160 | 912 | 2520 |
| **EvolveGCN** | 142,488 | 31,210 | 88,930 | 496.50 | 1952 |
| | **52%** | **22%** | **16%** | **54%** | **77%** |
| **GCRN-M2** | 151,302 | 27,482 | 121,088 | 382.50 | 2242 |
| | **55%** | **19%** | **22%** | **42%** | **89%** |

TABLE III: Information of datasets used in the experiments.

| Dataset | Avg nodes | Avg edges | Max nodes | Max edges | Time splitter | Snapshot count |
|---|---|---|---|---|---|---|
| **BC-Alpha** | 107 | 232 | 578 | 1686 | 3 weeks | 137 |
| **UCI** | 118 | 269 | 501 | 1534 | 1 day | 192 |

### B. Dataset Information

We use two publicly available benchmark dynamic graph datasets in the experiment. Bitcoin Alpha (BC-Alpha) [28] contains information related to user–user trust/distrust network from the Bitcoin Alpha trading platform. A node represents a trader and an edge represent the level of trust between traders. UC Irvine messages (UCI) [29] contains information about an online community of students from the University of California, Irvine. A node represents a user and an edge represents a sent message. More details about the datasets are shown in Table III.

### C. End-to-end Evaluation

We fully evaluate EvolveGCN and GCRN-M2 after place-and-route against CPU and GPU baselines. We measure latency on-board end-to-end, including weight loading and graph loading, and average across the snapshots in each dataset.

Results are depicted in Table IV. It shows that, for both EvolveGCN and GCRN-M2, DGNN-Booster achieves remarkable speedup over CPU and GPU. Since the message passing mechanism is not hardware-friendly to GPU [30] and also temporal data dependencies and frequent data exchange cause low GPU resource utilization and a large communication overhead between CPU and GPU [31], the latency reported by GPU baseline is a little higher than CPU.

TABLE IV: On-board latency (in ms) of EvolveGCN and GCRN-M2 per snapshot.

| Model (Dataset) | CPU | GPU | FPGA (Ours) | FPGA (vs. CPU) | FPGA (vs. GPU) |
|---|---|---|---|---|---|
| EvolveGCN (BC-Alpha) | 3.18 | 4.01 | 0.76 | **4.16×** | **5.25×** |
| EvolveGCN (UCI) | 3.68 | 4.19 | 0.86 | **4.26×** | **4.84×** |
| GCRN-M2 (BC-Alpha) | 7.39 | 11.35 | 1.35 | **5.48×** | **8.42×** |
| GCRN-M2 (UCI) | 8.50 | 9.74 | 1.51 | **5.63×** | **6.45×** |

TABLE V: Energy efficiency (including idle and runtime in J/100 snapshots) of DGNN-Booster.

| Model (Dataset) | CPU | GPU | FPGA (Ours) | FPGA (vs. CPU) | FPGA (vs. GPU) |
|---|---|---|---|---|---|
| EvolveGCN (BC-Alpha) | 5.84 | 32.16 | 1.92 | **3.04×** | **16.75×** |
| EvolveGCN (UCI) | 6.64 | 32.97 | 2.13 | **3.12×** | **15.48×** |
| GCRN-M2 (BC-Alpha) | 15.29 | 73.03 | 3.17 | **4.82×** | **23.04×** |
| GCRN-M2 (UCI) | 17.59 | 85.14 | 3.54 | **4.97×** | **24.05×** |

TABLE VI: Runtime energy efficiency (in J/100 snapshots) of DGNN-Booster.

| Model (Dataset) | CPU | GPU | FPGA (Ours) | FPGA (vs. CPU) | FPGA (vs. GPU) |
|---|---|---|---|---|---|
| EvolveGCN (BC-Alpha) | 1.83 | 21.01 | 0.02 | **91.5×** | **1050.5×** |
| EvolveGCN (UCI) | 2.08 | 21.54 | 0.03 | **69.3×** | **718×** |
| GCRN-M2 (BC-Alpha) | 6.57 | 47.71 | 0.05 | **131.4×** | **954.2×** |
| GCRN-M2 (UCI) | 7.56 | 55.63 | 0.06 | **126.0×** | **927.2×** |

Additionally, we evaluate the energy efficiency of DGNN-Booster on BC-Alpha and UCI datasets using a power meter connected to the board, as shown in Fig. 5. Table V and VI show the total and runtime energy efficiency of DGNN-Booster compared to CPU and GPU baselines respectively. The remarkable speedup and energy efficiency demonstrate the effectiveness of DGNN-Booster.

*D. Design space exploration and ablation study*

We perform a simple design space exploration to select the optimal accelerator configuration, taking into account the trade-off between GNN and RNN computations. In DGNN-Booster V1, we allocate more DSPs to RNN since it is computationally heavier than GNN. Conversely, in DGNN-Booster V2, we allocate more DSPs to GNN since GNN is computationally heavier. Table VII provides a detailed breakdown of the DSP allocation for each version. We further evaluate DGNN-Booster through an ablation study by comparing it to a GPU and a non-optimized FPGA baseline. We test two levels of pipeline optimization (Pipeline-O1 and Pipeline-O2). Pipeline-O1 involves pipelining different stages inside RNN, Pipeline-O2 adds more optimizations by overlapping GNN and RNN at a module level. Fig. 6 shows the incremental improvement of inference speed over the two baselines. It demonstrates that pipelining GNN and RNN from a multi-level can be very effective in reducing latency.

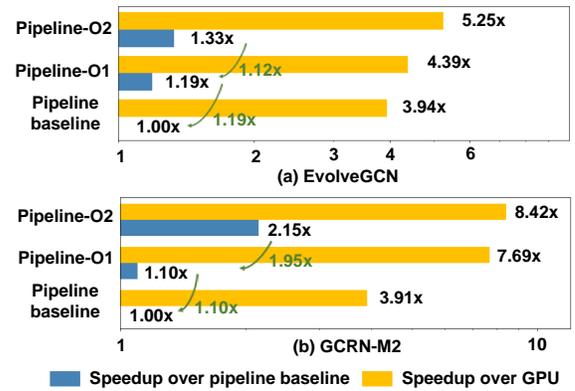

Fig. 6: Effectiveness of DGNN-Booster's two designs. Speedup plotted in log-scale. **Baseline:** Without optimizations proposed in this paper. **Pipeline-O1:** Pipelines different stages inside RNN. **Pipeline-O2:** Overlaps GNN and RNN.

TABLE VII: Design space exploration results. The latency is averaged from all of the snapshots in BC-Alpha and UCI datasets. DSP utilization results are reported by Vivado.

| Framework | Module | Latency (ms) | DSP |
|---|---|---|---|
| **DGNN-Booster V1** (EvolveGCN) | GNN | 0.36 (43%) | 288 (15%) |
| | RNN | 0.47 (57%) | 1658 (85%) |
| **DGNN-Booster V2** (GCRN-M2) | GNN | 0.82 (49%) | 2171 (96%) |
| | RNN | 0.85 (51%) | 78 (4%) |

## VI. CONCLUSIONS AND FUTURE WORKS

In this paper, we propose **DGNN-Booster**, which is an FPGA accelerator framework including two designs with different dataflow patterns and multi-level parallelism. DGNN-Booster is open-sourced and can achieve real-time inference speed with low energy consumption. Moreover, it is a generic framework that supports various types of GNNs and RNNs, integrated with GNN and RNN hardware optimizations together to achieve optimal performance improvement. Future works include the on-board implementation to avoid redundant data communication and computation because of the similarity between snapshots in adjacent time steps. We also plan to do design space exploration to balance computation resources for spatial and temporal encoding parts.

ACKNOWLEDGMENT

This work is partially supported by National Science Foundation (NSF) Grant ECCS-2202329.